\documentstyle[aps,leqno,preprint]{revtex}  
\input{epsf}
\begin{document}
\renewcommand{\thefootnote}{\alph{footnote} }
\narrowtext
%
%
\hoffset=0.0in
%
\def\DZ{\ifmmode D^0 \else $D^0$\fi}
\def\DZb{\ifmmode \bar{D}^0 \else $\bar{D}^0$\fi}
\def\DP{\ifmmode D^+ \else $D^+$\fi}
\def\DSP{\ifmmode D^{*+} \else $D^{*+}$\fi}
\def\DSM{\ifmmode D^{*+} \else $D^{*-}$\fi}
\def\DST{\ifmmode D^* \else $D^*$\fi}
\def\DM{\ifmmode D^- \else $D^-$\fi}
\def\DPM{\ifmmode D^\pm \else $D^\pm$\fi}
\def\Mmin{\ifmmode M_{\rm min} \else $M_{\rm min}$\fi}
\def\PT{\ifmmode p_T \else $p_T$\fi}
\def\KM{\ifmmode K^- \else $K^-$\fi}
\def\kstar{\ifmmode K^{*0} \else $K^{*0}$\fi}
\def\kstarb{\ifmmode \overline{K}^{*0} \else $\overline{K}^{*0}$\fi}
\def\kstarbno{\ifmmode \overline{K}^{*} \else $\overline{K}^{*}$\fi}
\def\kst892{\ifmmode \overline{K}^{*}(892)^0 \else 
$\overline{K}^{*}(892)^0$\fi}
\def\kstz{\ifmmode \overline{K}^{*}(892)^{0} 
  \else $\overline{K}^{*}(892)^{0}$\fi}
\def\KSTM{\ifmmode \overline{K}^{*-} \else $\overline{K}^{*-}$\fi}
\def\kstm890{\ifmmode \overline{K}^{*}(892)^- 
\else $\overline{K}^{*}(892)^-$\fi}
\def\kst{\ifmmode \overline{K}^{*}(892) \else 
$\overline{K}^{*}(892)$\fi}
\def\rh{\ifmmode \rho(770) \else $\rho(770)$\fi}
\def\rh0{\ifmmode \rho(770)^{0} \else $\rho(770)^{0}$\fi}
\def\rhm{\ifmmode \rho(770)^{-} \else $\rho(770)^{-}$\fi}
\def\munu{\ifmmode \mu^{+}\nu \else $\mu^{+}\nu$\fi}
\def\muX{\ifmmode \mu^{+}X \else $\mu^{+}X$\fi}
\def\lnu{\ifmmode \ell\nu \else $\ell\nu$\fi}
\def\nmu{\ifmmode \nu_\mu \else $\nu_\mu$\fi}
\def\nmub{\ifmmode \bar{\nu}_\mu \else $\bar{\nu}_\mu$\fi}
\def\pmu{\ifmmode p_\mu \else $p_\mu$\fi}
\def\ptmu{\ifmmode p_{T\mu} \else $p_{T\mu}$\fi}
\def\pth{\ifmmode p_{T h} \else $p_{T h}$\fi}
\def\bmu{\ifmmode b_\mu \else $b_\mu$\fi}
\def\pperp{\ifmmode p_\perp \else $p_\perp$\fi}
\def\xf{\ifmmode x_F \else $x_F$\fi}
\def\pts{\ifmmode {p_T}^2 \else ${p_T}^2$\fi}
\def\pt{\ifmmode {p_T} \else ${p_T}$\fi}
\def\chisq{\ifmmode \chi^2 \else $\chi^2$\fi}

\def\pvec{\ifmmode \underline{p} \else $\underline{p}$\fi} 
%
\def\sameauthors#1
{\hbox to\textwidth{\hss\vrule height.3cm width0pt\relax%
#1\hss}}
\def\etal{{\it et al.}\rm}
%

\title{{\bf 
Asymmetries between the production of $D^+$ and $D^-$ mesons  from 500 
GeV$/c$  $\pi^-$-nucleon interactions as a function of
$x_F$ and $p_t^2$  \\}}

%
%
%
\author{
    E.~M.~Aitala,$^8$
       S.~Amato,$^1$
    J.~C.~Anjos,$^1$
    J.~A.~Appel,$^5$
       D.~Ashery,$^{14}$
       S.~Banerjee,$^5$
       I.~Bediaga,$^1$
       G.~Blaylock,$^2$
    S.~B.~Bracker,$^{15}$
    P.~R.~Burchat,$^{13}$
    R.~A.~Burnstein,$^6$
       T.~Carter,$^5$
 H.~S.~Carvalho,$^{1}$
       I.~Costa,$^1$
    L.~M.~Cremaldi,$^8$
 C.~Darling,$^{18}$
       K.~Denisenko,$^5$
       A.~Fernandez,$^{11}$
       P.~Gagnon,$^2$
       S.~Gerzon,$^{14}$
       C.~Gobel,$^1$
       K.~Gounder,$^8$
       D.~Granite,$^7$
     A.~M.~Halling,$^5$
       G.~Herrera,$^4$
 G.~Hurvits,$^{14}$
       C.~James,$^5$
    P.~A.~Kasper,$^6$
 N.~Kondakis,$^{10}$
       S.~Kwan,$^5$
    D.~C.~Langs,$^{10}$
       J.~Leslie,$^2$
       J.~Lichtenstadt,$^{14}$
       B.~Lundberg,$^5$
       A.~Manacero,$^5$
       S.~MayTal-Beck,$^{14}$
       B.~Meadows,$^3$
 J.~R.~T.~de~Mello~Neto,$^1$
    R.~H.~Milburn,$^{16}$
 J.~M.~de~Miranda,$^1$
       A.~Napier,$^{16}$
       A.~Nguyen,$^7$
  A.~B.~d'Oliveira,$^{3,11}$
       K.~O'Shaughnessy,$^2$
    K.~C.~Peng,$^6$
    L.~P.~Perera,$^3$
    M.~V.~Purohit,$^{12}$
       B.~Quinn,$^8$
       S.~Radeztsky,$^{17}$
       A.~Rafatian,$^8$
    N.~W.~Reay,$^7$
    J.~J.~Reidy,$^8$
    A.~C. Reis,$^1$
    H.~A.~Rubin,$^6$
 A.~K.~S.~Santha,$^3$
 A.~F.~S.~Santoro,$^1$
       A.~J.~Schwartz,$^{10}$
       M.~Sheaff,$^{17}$
    R.~A.~Sidwell,$^7$
    A.~J.~Slaughter,$^{18}$
    J.~G.~Smith,$^7$
    M.~D.~Sokoloff,$^3$
       N.~R.~Stanton,$^7$
       K.~Sugano,$^2$
    D.~J.~Summers,$^8$
 S.~Takach,$^{18}$
       K.~Thorne,$^5$
    A.~K.~Tripathi,$^9$
       S.~Watanabe,$^{17}$
 R.~Weiss-Babai,$^{14}$
       J.~Wiener,$^{10}$
       N.~Witchey,$^7$
       E.~Wolin,$^{18}$
       D.~Yi,$^8$
       R.~Zaliznyak,$^{13}$
       and
       C.~Zhang$^7$ \\
(Fermilab E791 Collaboration) 
}


\address{
$^1$ Centro Brasileiro de Pesquisas F\'{\i}sicas, Rio de Janeiro, Brazil\\
$^2$ University of California, Santa Cruz, California 95064\\
$^3$ University of Cincinnati, Cincinnati, Ohio 45221\\
$^4$ CINVESTAV, Mexico\\
$^5$ Fermilab, Batavia, Illinois 60510\\
$^6$ Illinois Institute of Technology, Chicago, Illinois 60616\\
$^7$ Kansas State University, Manhattan, Kansas 66506\\
$^8$ University of Mississippi, University, Mississippi 38677\\
$^9$ The Ohio State University, Columbus, Ohio 43210\\
$^{10}$ Princeton University, Princeton, New Jersey 08544\\
$^{11}$ Universidad Autonoma de Puebla, Mexico\\
$^{12}$ University of South Carolina, Columbia, South Carolina 29208\\
$^{13}$ Stanford University, Stanford, California 94305\\
$^{14}$ Tel Aviv University, Tel Aviv, Israel\\
$^{15}$ 317 Belsize Drive, Toronto, Canada\\
$^{16}$ Tufts University, Medford, Massachusetts 02155\\
$^{17}$ University of Wisconsin, Madison, Wisconsin 53706\\
$^{18}$ Yale University, New Haven, Connecticut 06511\\
}

\rm

\date{\today}

\maketitle

\begin{abstract}
  We present asymmetries between the production of 
$D^+$ and $D^-$ mesons  in Fermilab experiment E791 
as a function of $x_F$ and $p_t^2$. The data used here
consist of   74,000  fully-reconstructed charmed  mesons produced by a 500 
GeV/$c$ $\pi^-$ beam on C and Pt foils. The measurements are compared 
to  results of models  which predict  differences between the production 
 of heavy-quark mesons that have a light quark in common 
with the beam (leading particles) and those that do not (non-leading 
particles). While the default models  do not agree with our data, we can 
reach agreement with one of them, PYTHIA, by making a limited number of 
changes to  parameters used.

\end{abstract}


Among the least understood aspects of quantum chromodynamics (QCD) is 
fragmentation, the
non-perturbative dressing of bare quarks into hadrons.
Certain asymmetries in the production of particles and antiparticles are
especially sensitive to fragmentation effects, while largely
free of experimental bias in the measurement process.
In this paper we present high statistics measurements of charmed particle
asymmetries from $\pi$-nucleon interactions at 
experiment E791 at Fermilab.

   Early studies of charm particle production \cite{exp:isr,exp:bas,exp:rit}
showed evidence of a large enhancement  in the forward production  of charmed
particles that contain  a quark or antiquark  in common with the beam 
(leading particles) over those that do not (non-leading particles).
Leading-order QCD calculations predict no asymmetry between quark and 
antiquark production.
Several other models were developed to explain this 
effect \cite{thr:hal,thr:maz}, although
none fully  accounted for the difference in production between
 leading and non-leading particles. As sample sizes and signal-to-background
ratios increased, experiments \cite{exp:wa82,exp:e769} with 
signals on the order of a thousand charmed mesons continued to find
a forward production asymmetry, one
much higher than
predicted even by next-to-leading-order (NLO) QCD calculations
which became available \cite{thr:nason}.
Now, we report results based on 74,000 fully 
reconstructed $D^\pm$  decays, enough to examine the asymmetry both as a function of 
Feynman-$x$ ($x_F \equiv p_z/p_z^{max})$ and transverse
momentum squared ($p_t^2$).

For a $\pi^-$ beam, NLO QCD calculations  predict
a  small enhancement in the number  of $\bar{c}$ quarks over $c$ quarks 
in the very forward direction.  However,  as has been previously 
shown \cite{exp:e769},
this does not fully account for the   observed 
leading particle enhancement.
We  compare our results to two current models which predict a much
larger enhancement than NLO QCD. We also report changes to parameter values  
within  one of these models (PYTHIA) \cite{thr:pyth,thr:pythc} that
provide  agreement with our data.

   One  possible explanation  of the asymmetry is given  by the 
beam-dragging model in which forward momentum is added to the 
produced heavy quark if it combines with a remnant
light quark from the incoming beam particle, forming a leading particle.
 This causes leading charmed
particles to have a harder $x_F$ spectrum than non-leading particles.  
This feature is characteristic of 
the Lund string-fragmentation model which is the basis
of the fragmentation simulation software JETSET used in PYTHIA. 
  
   A  second class of  models\cite{thr:brod,thr:wa,thr:olga}
postulates charm may be produced as virtual $c\bar{c}$ pairs in the  
beam particle in addition to being produced perturbatively.  
An example of the former  process is the intrinsic charm model
proposed by Vogt and Brodsky \cite{thr:brod}.
Here, the incoming $\pi^-$ beam particle fluctuates into a
$|\bar{u}dc\bar{c}~\rangle$ Fock state.
Since the virtual   $c\bar{c}$ quarks have nearly the  same 
velocity as the original $\bar{u}$
and $d$ quarks in the pion, they are likely to coalesce with these 
remnant quarks,  forming  leading particles. Charmed quarks can also
hadronize via fragmentation mechanisms which do not distinguish between 
leading and non-leading charm production. In reference~\cite{thr:brod} 
specifically, the fragmentation mechanism assumes that the momentum of the 
$D$ meson is equal to the momentum of the $c$ quark.

   To quantify the difference in the production of leading and non-leading
particles, an asymmetry parameter $A$ is defined for each region
of phase space:

\begin{equation}
  A(x_F,p_t^2) \equiv \frac{\sigma_{L} - \sigma_{NL}}{\sigma_{L} + \sigma_{NL}}.
  \label{eq:aeq}
\end{equation}

\noindent{Here}, $\sigma_L(\sigma_{NL})$ is the production cross section for the 
leading (non-leading)  particle being studied.
Experiments WA82 \cite{exp:wa82}
and E769 \cite{exp:e769}
found large values of  $A$ ($\simeq 0.5)$
for high values of $x_F$ ($\simeq 0.6$).

   For the $\pi^- (d\bar{u})$ beam in E791, the $D^0(c\bar{u})$ and $D^-(\bar{c}d)$ are
leading charmed   mesons. However, a
large fraction of the $D^0$'s are  produced by the 
$D^{*+} \rightarrow D^0 \pi^+$ decay process.  The original $D^{*+}$ is
actually a {\em non-}leading particle.  Therefore, the observed $D^0$'s come
from a mixture of leading and non-leading processes, making their  study
more complex.  In order to minimize systematic errors, we restrict ourselves
to studies of the asymmetries in the $D^\pm$ spectra.

	The data were recorded from 500 GeV/$c$ $\pi^-$ interactions in five
thin foils (one platinum, four diamond) separated by gaps of 1.34 to
1.39 cm. The E791 spectrometer is an upgraded version of the apparatus used
in Fermilab experiments E516, E691, and E769 \cite{detector}. Momentum analysis
is provided by two dipole magnets which bend 
particles in the horizontal plane and 35 planes of drift chambers.
Silicon microstrip detectors (6 in the beam, 17 downstream of the target) provide
precision track and vertex reconstruction.  

	E791 recorded $2 \times 10^{10}$ events with a loose transverse energy trigger.
After reconstruction, events with evidence of multiple vertices are kept for
further analysis. We search for $D^\pm$ mesons in the 
decay mode $D^+(D^-)\rightarrow 
K^-\pi^+\pi^+ (K^+\pi^-\pi^-)$.
The selection criteria listed below are chosen to maximize
$N_S/\sqrt{N_S + N_B}$ where $N_S$ and $N_B$ are the number of signal and background
events.

\begin{figure}[htb]
\centerline {\epsfxsize=6.0in \epsffile{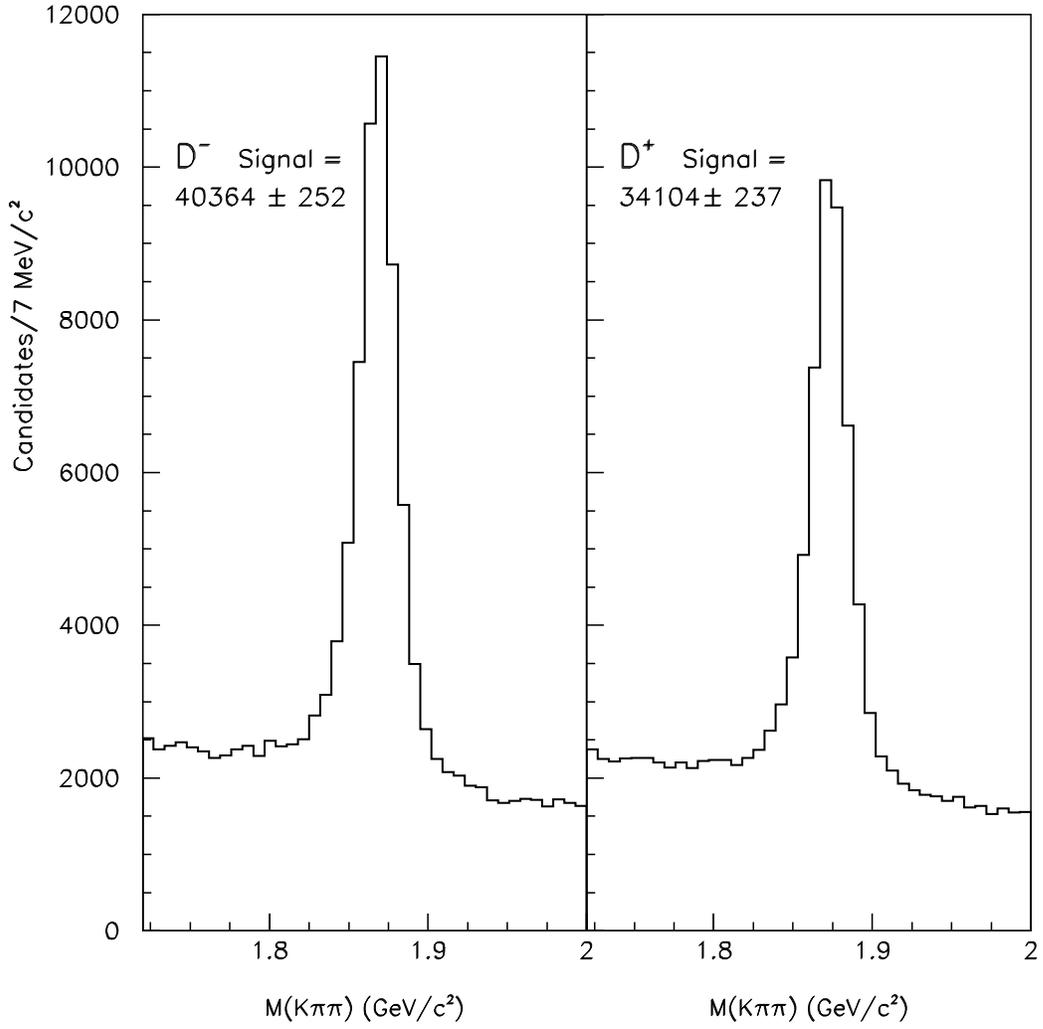} }
\caption{$K\pi\pi$ mass distributions for $D^+$ and $D^-$ candidates after
all selection criteria are applied.}
\label{fig:peaks}
\end{figure}

To help ensure that the reconstructed secondary vertex is a
true decay vertex,
we require that it be separated from the
primary vertex by at least 9.0 $\sigma_l$, where $\sigma_l$ is our resolution on
the separation, and separated from the nearest target
foil by at least 0.75 $\sigma_l$. In addition, we impose two requirements
to ensure that the candidate decay tracks do not originate
at the primary vertex. First,
each of the three decay tracks must have an impact
parameter with respect to the primary vertex greater than 
5.5 $\sigma_p$, where 
$\sigma_p$
is the resolution on the impact parameter. Second,
we form the ratio of each track's distance from the secondary
vertex to that from the primary vertex and require the product
of these three ratios to be less than $0.002$.
Another two requirements ensure that
the reconstructed
charged $D$ is consistent with 
originating at the primary vertex.
First, the impact parameter of the
reconstructed $D$ momentum vector with respect to the primary vertex
must be less than 50 $\mu m$ \cite{dip}.
Second, the component of the reconstructed
$D$ momentum perpendicular to the $D^\pm$ line-of-flight
(as determined from the primary and secondary vertex positions)
must be less than
0.7 GeV/$c$. Finally, to help ensure that the three measured
tracks come from the decay of a high mass particle,
the sum of the squares of the tracks'  momenta transverse to the direction 
of the reconstructed candidate momentum must be 
greater than 0.4 (GeV/$c)^2$. The resulting $K\pi\pi$ mass spectra are shown in 
Fig.~\ref{fig:peaks} for $D^+$ and $D^-$ candidates.

  The $K\pi\pi$ spectrum,
in each region of phase space,
was fit
to a Gaussian curve for the signal plus a linear distribution for the
background. 
The number of $D$ candidates is determined from
the total area under the Gaussian curve.
The width of the Gaussian was fixed for each region, with the 
evolution of the width as a function of $x_F$ determined 
by Monte Carlo simulations. 
The systematic error in the number of mesons caused by
the uncertainty  in the width was estimated by increasing and decreasing the width by $10\%$, recalculating 
the yield each time. The resulting systematic uncertainties are 
small compared to the statistical uncertainties.

 Detailed studies  of our charged particle reconstruction
 using both data and Monte Carlo
simulations show a small difference in reconstruction efficiency between 
positively and negatively charged particles
due mainly to degradation 
over time of the wire tracking chambers in the vicinity of the incident
beam.  Because the
negative beam particles are bent by the analysis magnets, the
center of  degradation was shifted slightly  from the center of
our detector. The effect
on the measured values of $A$ was determined using Monte Carlo
simulations and is significant only in the high $x_F$ region, increasing the value of
$A$ by $\sim 0.15$ in the highest bin.  The systematic uncertainty 
caused by the efficiency correction was estimated by modeling the degradation of our
chambers for four different time periods
during our run, including periods near the  beginning and the
end of our data-taking.  
The rms spread of the asymmetries from the four
different efficiency corrections was used to estimate 
the systematic uncertainties.
These uncertainties are comparable to the uncertainties resulting from
statistical errors.

\begin{figure}[htb]
\centerline {\epsfxsize=6.0in \epsffile{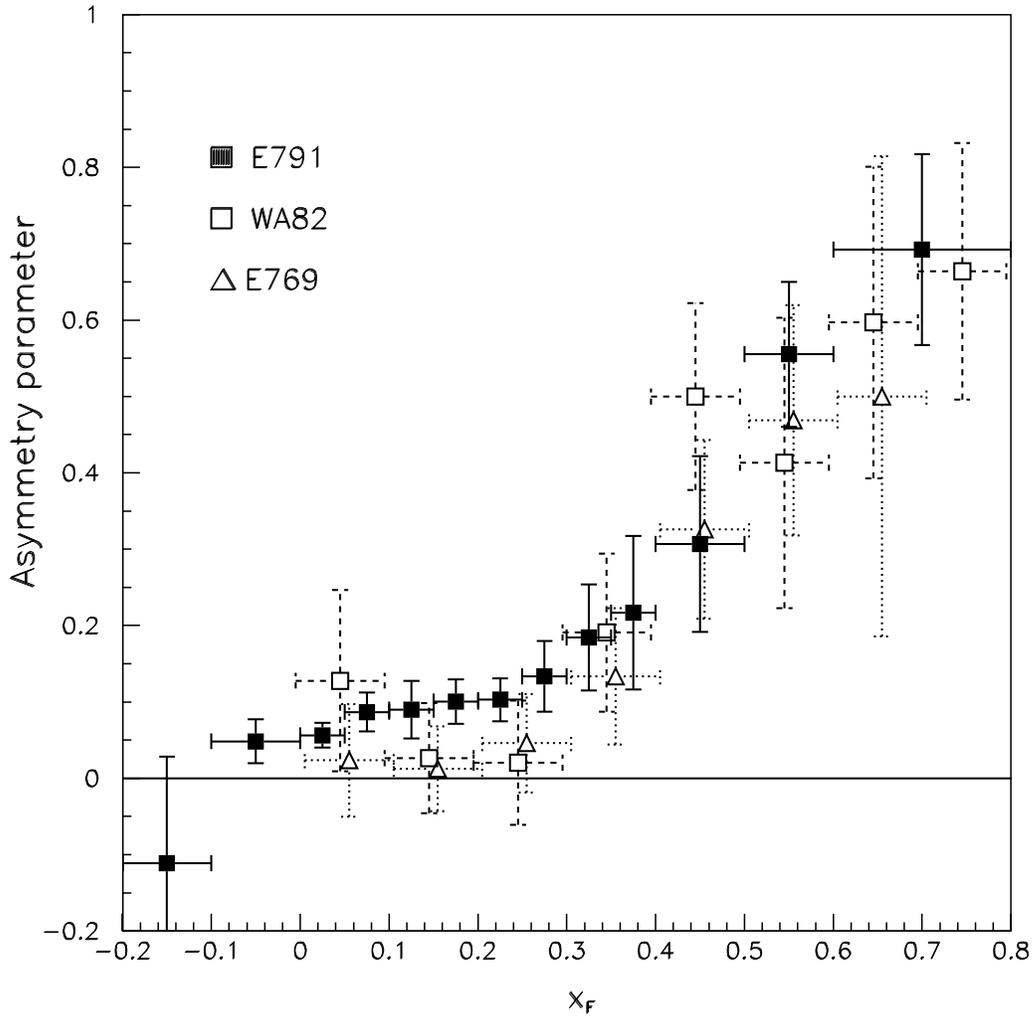} }
\caption{The charged $D$ production
 asymmetry $A$ as a function of $x_F$.
The E791 results are compared to previous results from 
WA82 [6]  and E769 [7]. WA82 data are for $D^\pm$ from a 340 GeV/$c$
$\pi^-$ beam, E769 data
are for $D^\pm$ and $D^{*\pm}$ from a 250 GeV/$c$ $\pi^\pm$ beam, and 
E791 data are for $D^\pm$ from a 500 GeV/$c$
$\pi^-$ beam.  E791 and E769 data are for mesons 
with  $p_t^2$ less than
10 (GeV/$c)^2$.}
\label{fig:axf_others}
\end{figure}

  Figure~\ref{fig:axf_others} shows the value of the asymmetry parameter 
calculated for different bins of $x_F$, compared to results from 
experiments E769 and WA82.
The E791 error
bars shown on all plots include both systematic 
errors (from the uncertainty on the mass resolution
and from the uncertainty on the reconstruction efficiency)
and statistical errors, added in quadrature.
Our measured asymmetries are higher than in those two experiments,  
especially at low $x_F$, perhaps due to the different beam
momenta (500 GeV$/c$ for E791, 
340 GeV$/c$ for WA82 and 250 GeV$/c$ for E769).

\begin{figure}[htb]
\centerline {\epsfxsize=6.0in \epsffile{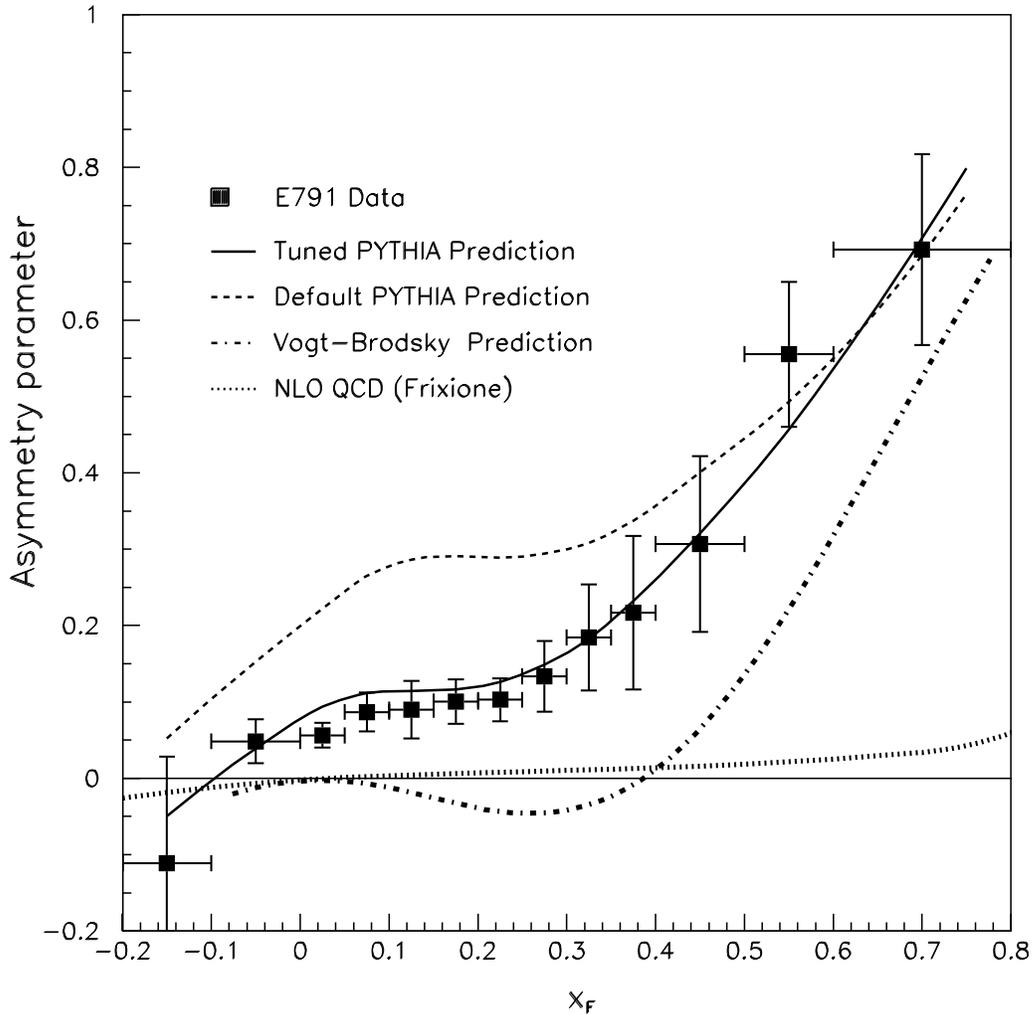} }
\caption{
The $D^\pm$ asymmetry $A$ as a function of $x_F$ for E791 data (points
with error bars), for three models, and for NLO QCD.
The tuned PYTHIA
prediction [16] is from the default PYTHIA Monte 
Carlo software with the $c$ quark mass increased
to 1.7 GeV/$c^2$ and the 
average primordial $k_t^2$ of partons increased to (1.0 GeV/$c)^2$, as
discussed in the text.  The Vogt-Brodsky prediction is specifically 
calculated for the  E791 beam momentum [18]. Both data and  model 
predictions are for $D^\pm$ mesons with $p_t^2$ less than 10 (GeV/$c)^2$.  
The NLO QCD prediction is for charm quarks rather than
$D$ mesons [8].}
\label{fig:axf_models}
\end{figure}

In Fig.~\ref{fig:axf_models} we compare our values of $A$ to predictions
based on two models for $D$ production previously discussed and 
to the prediction for charm quarks by NLO QCD.
The results  from the Lund string fragmentation model used in PYTHIA
are shown in two different curves.  The dashed line shows the
prediction from the default PYTHIA and is significantly higher than our
data for $-0.2 < x_F < 0.4$. This is due to the fact that 
PYTHIA predicts a higher overall production ratio  
of $D^-$ to $D^+$ than is seen in data.  
We also  explored a range of parameters to determine if our results could be
accommodated within the PYTHIA model.
A tuned PYTHIA  prediction \cite{cparam} 
that is  consistent with our data 
is shown as the solid line in Fig.~\ref{fig:axf_models}.
This tuning included increasing the 
mass of the $c$ quark ($m_c$)  from 1.35 GeV$/c^2$ to 
1.7 GeV/$c^2$ and increasing the average primordial $k_t^2$ of the partons from (0.44 
GeV/$c)^2$ to (1.0 GeV/$c)^2$.  This decreases the likelihood that the remnant $d$ quark can
combine with the $\bar{c}$ quark with a small enough invariant mass to collapse to a $D^-$ 
meson.  
  While this value of mean $k_t^2$ may seem
unphysically high, 
similar values of $k_t^2$ are suggested by other observations from 
photo- and hadro-production 
charm experiments \cite{thr:frix}.

In  Fig.~\ref{fig:axf_models} we also compare our results to a recent prediction involving
intrinsic charm by Vogt and Brodsky, specifically  calculated for a  500 GeV/{\em c} $\pi^-$ beam 
\cite{thr:brod,priv:vogt}.
Although the shape of the $A$ vs $x_F$ curve is similar to our data,
the prediction is too low for all $x_F$.
This model assumes equal numbers of $D^-$ and $D^+$ mesons 
were produced, unlike the PYTHIA model, and  changing   this ratio may produce
better agreement with data.

\begin{figure}[htb]
\centerline {\epsfxsize=6.0in \epsffile{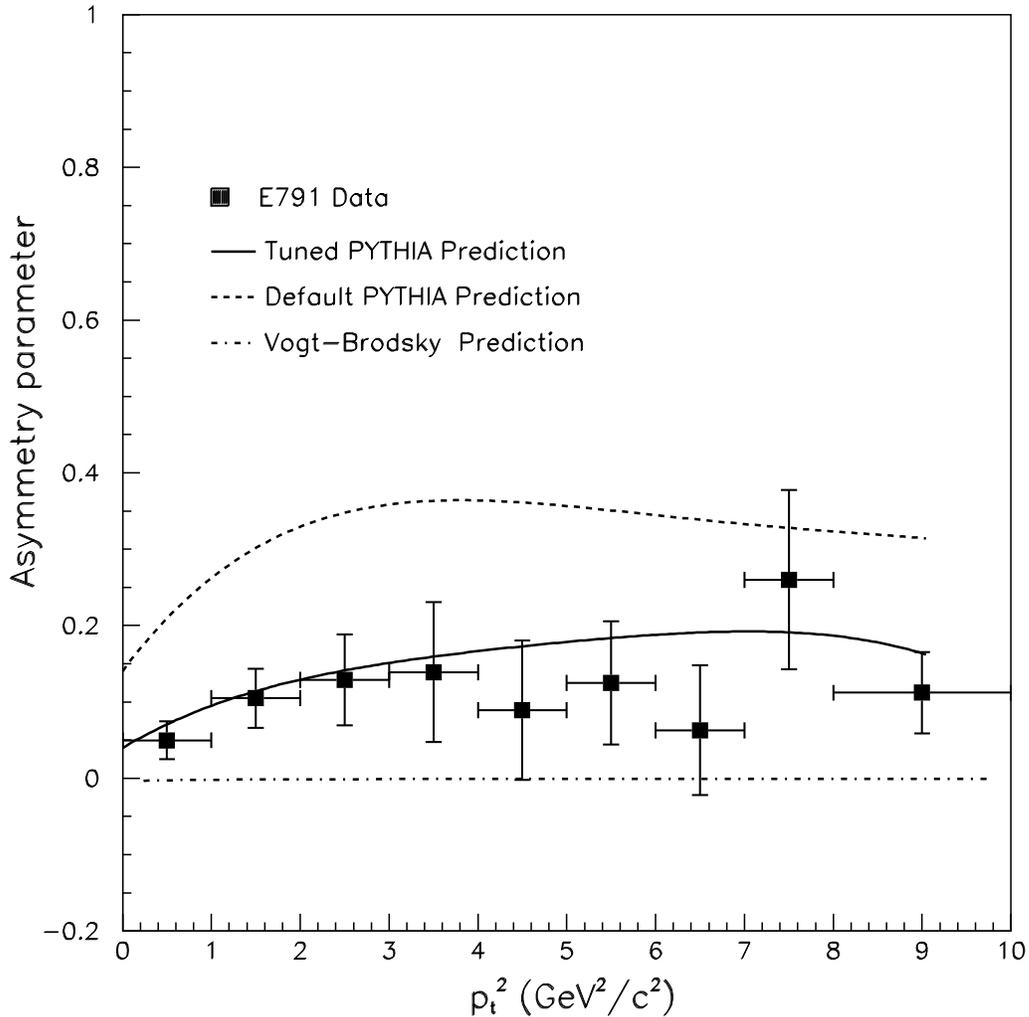} }
\caption{
The $D^\pm$ asymmetry $A$ as a function of $p_t^2$ for our full
range of $x_F$
($-0.2 \leq x_F \leq 0.8$). The data points correspond to the measured
values. The curves represent theoretical predictions.
The tuned PYTHIA
prediction [16] is from the default PYTHIA Monte 
Carlo software with the $c$ quark mass increased
to 1.7 GeV/$c^2$ and the average primordial $k_t^2$ of partons increased 
to (1.0 GeV/$c)^2$ as discussed in the text.  The intrinsic charm 
prediction is specifically calculated for the 
E791 beam momentum [18]. }
\label{fig:apt2_models_all}
\end{figure}

   Figure~\ref{fig:apt2_models_all} shows our data compared to the predictions for $A$ 
from these same models versus $p_t^2$ for
$-0.2 \leq x_F \leq 0.8 $. Again, the default PYTHIA 
model predicts values which are too high for most points, while the
tuned PYTHIA is in good agreement with our data. The intrinsic charm model of Vogt
and Brodsky predicts values
close to zero,  indicative  of the assumption that $D^-$ and $D^+$ mesons are produced
in equal numbers.

\begin{figure}[htb]
\centerline {\epsfxsize=6.0in \epsffile{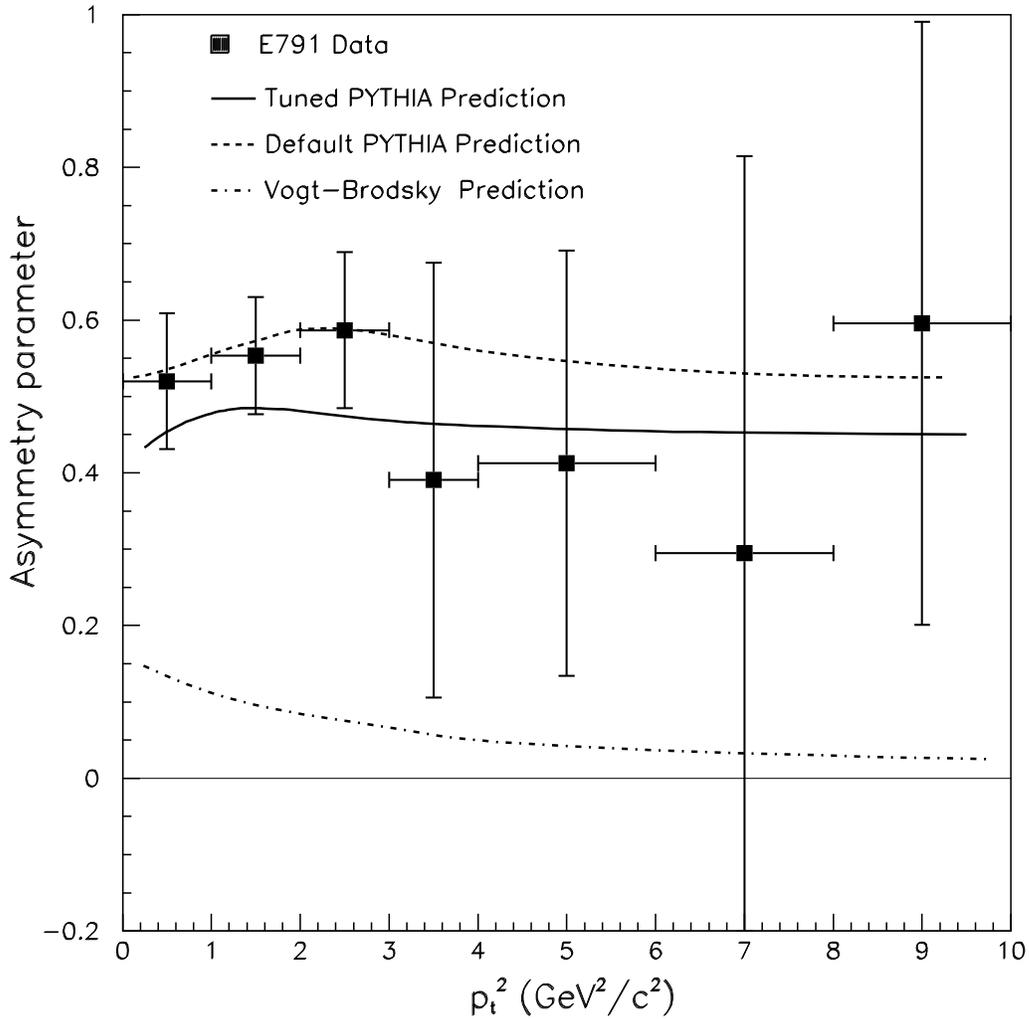} }
\caption{
The $D^\pm$ asymmetry $A$ as a function of $p_t^2$ for high values
of $x_F$
($0.4 \leq x_F \leq 0.8$). The data points correspond to the measured
values. The curves represent theoretical predictions. The rise in $A$
at low values of $p_t^2$ predicted by the Vogt-Brodsky model is
a key feature which is not reproduced by
our data.
}
\label{fig:apt2_models_hi}
\end{figure}

In the intrinsic charm model, coalescence occurs dominantly at low $p_t^2$
$(\sim m_c^2)$ \cite{thr:brod}.  Thus, in the context of coalescence alone,
$A$ should be large at high $x_F$ and low $p_t^2$, increasing as $p_t^2$ goes
to zero. In order to address this issue, we show the asymmetry as a function
of $p_t^2$ for mesons with high $x_F$ values ($0.4 \leq x_F \leq 0.8$)
in Fig.~\ref{fig:apt2_models_hi}.  Here, the intrinsic charm model predicts 
values of $A$ that are too low to match our data. However, we note that 
fragmentation mechanisms alternative to the one used in reference [11] can 
soften equally the $x_F$ spectrum of both the leading and non-leading $D$'s.
This would emphasize the hardness of the $x_F$ spectrum of 
leading $D$'s produced by coalescence and thereby increase the predicted
asymmetry at high $x_F$. Independent of this over all scale, the 
model predicts an increase in   the value of $A$ as $p_t^2$ goes to zero.
Our data show no indication of this increase.

  In summary, we see a substantial  leading-particle effect at high $x_F$ which cannot
be accounted for using current  perturbative QCD calculations. The  beam-dragging model used in PYTHIA 
can account for the asymmetries
between $D^-$ and $D^+$ production if reasonable changes are 
made to the parameters used in the model.  Neither the default PYTHIA nor  
the intrinsic charm model is  consistent with our data.


We gratefully acknowledge the funding from the U.S. Department of
Energy, the U.S. National Science Foundation, the Brazilian Conselho
Nacional de Desenvolvimento Cient\'{\i}fico e Tecnol\'ogico, the Mexican 
CONACyT, the Israeli 
Academy of Science and the U.S.-Israeli Binational Science Foundation.  We would
like to thank experiment E769 for providing us with detailed information on their
results and T. Sjostrand, M. Mangano, S. Frixione, R. Vogt, S.J. Brodsky, and  O. Piskounova
for discussions.

\bibliographystyle{unsrt}

\begin{thebibliography}{99.}
\bibitem{dip}
The precise requirement we impose is:
$d \leq (\Delta_l/\sigma_l)(0.002$ cm), where $d$ is
the $D$ impact parameter,
$\Delta_l$ is the separation of the
$D$ vertex from the nearest target foil and
$\sigma_l$ is the resolution on the separation.
In addition, the maximum allowed value for $d$ is 0.005~{\ cm}.
\bibitem {cparam} Specific parameters that we changed were {\tt PMAS(4,1)}
(the $c$ quark mass in GeV/$c^2$) increased from {\tt 1.35} to {\tt 1.7}, 
{\tt PARP(91)}  (the average $k_t^2$ in (GeV/$c)^2$) increased from {\tt 0.44}
to {\tt 1.0}, {\tt PARP(93)} (the maximum allowable $k_t$ in GeV/$c$) 
increased from {\tt 2.0} to {\tt 5.0}, and {\tt MSTP(92)} (the remnant 
quark-diquark energy splitting function) changed from {\tt 4} to {\tt 3}.
\end{thebibliography}

\newpage

\newpage


\newpage

\newpage

\newpage

\end{document}